\documentclass[12pt,english]{article}
\usepackage[T1]{fontenc}
\usepackage[latin9]{inputenc}
\usepackage{amstext}
\usepackage{graphicx}
\usepackage{esint}

\makeatletter

\providecommand{\tabularnewline}{\\}

\newcommand{\lyxaddress}[1]{
\par {\raggedright #1
\vspace{1.4em}
\noindent\par}
}

\@ifundefined{showcaptionsetup}{}{%
 \PassOptionsToPackage{caption=false}{subfig}}
\usepackage{subfig}
\makeatother

\usepackage{babel}
\begin{document}

\title{Analysis of High Power Behavior in Piezoelectric Ceramics from a
Mechanical Energy Density Perspective}

\author{H. N. Shekhani$^{1}$, E. A. Gurdal$^{2}$, S. O. Ural$^{3}$, and
K. Uchino$^{1}$}

\maketitle

\lyxaddress{$^{1}$International Center for Actuators and Transducers, Materials
Research Institute, The Pennsylvania State University, University
Park, PA, USA\\$^{2}$Center for Dielectrics and Piezoelectrics,
Materials Research Institute, The Pennsylvania State University, University
Park, PA, USA\\$^{3}$Morgan Advanced Materials, Cleveland, Ohio,
USA}
\begin{abstract}
In this work, a new methodology for comparing high power performance
of different piezoelectric materials is presented. When comparing
high power performance of piezoelectric materials of varying compositions
and vibration modes, there exists an inherent problem in comparing
the mechanical quality factor directly. Typically the behavior of
the mechanical quality factor is reported as a function of tip vibration
velocity of the sample. This paper shows why this approach can be
inherently problematic and proposes an energy approach to characterize
the mechanical quality factor as the solution. By utilizing mechanical
energy density ($u_{e}$), the mass density of the material system
($\rho$), and the vibration mode shape (e.g. $k_{31}$ and $k_{p}$)
of the sample are accounted for. Therefore, a better method to compare
high power performance of varying piezoelectric compositions is introduced.
Furthermore, the new method is applied to various compositions by
using data available in the literature. As a result, the high power
behavior of the materials appear to be significantly different when
the new ``energy density approach'' is used to compare the mechanical
quality factor rather than vibration velocity. Also, the technique's
ability to be utilized to consolidate data from different resonators
to determine anisotropic loss factors is demonstrated on hard and
soft PZT samples of $k_{31}$ and $k_{p}$ geometries.
\end{abstract}

\section{Introduction}

The measured properties of piezoelectric materials are sensitive to
the conditions under which they are tested. Traditionally, piezoelectric
materials are characterized under low amplitude excitation conditions,
which provide different properties than that of large excitation,
whose practical applications include popular transducers such as piezoelectric
transformers and ultrasonic motors \cite{uchino2009ferroelectric}.
Therefore, many researchers have taken to study the high power characteristics
of piezoelectric materials because the properties at high power levels
are more relevant from an application standpoint. \cite{Umeda1998,Hirose1996,Uchino1998}

The mechanical quality factor ($Q_{m}$), is among the most critical
properties for the design and application of high power piezoelectric
transducers. The mechanical quality factor provides an amplification
factor for strain and vibration in resonance conditions. The inverse
of the mechanical quality factor provides a measure of the energy
lost per cycle relative to the stored energy due to the hysteresis.
Thus, the $Q_{m}$ is a very important parameter to discuss when evaluating
a material's high power performance. The high quality factor, or the
``hardness'' of the industry standard piezoelectric material, lead
zirconate titanate (PZT), is achieved by acceptor doping the system
and thereby generating internal field to suppress domain dynamics,
the largest source of loss in this material system. With the recent
environmental concerns due to lead content in PZT and other lead containing
piezoelectric materials, many researchers are searching for lead free
piezoelectric materials with a large mechanical quality factor to
replace hard PZT for high power applications \cite{Uchino2003}. 

Many researchers have produced various promising materials which might
replace PZT and other lead containing materials in high power applications\cite{Takenaka1991,Shrout2007}.
One of the objectives of this paper is to clarify a better method
to compare the performance of different piezoelectric material systems
in high power conditions. In order to compare materials having different
sample dimensions, the vibration velocity condition has been used.
This methodology is based on the fact that the strain levels between
samples with different dimensions are the same if they are under the
same vibration velocity condition. Thus, the vibration condition can
be used as a common excitation condition to compare results\cite{Uchino2000}.
This, however, only applies to samples having the same mode of vibration.
For example, the radial vibration mode of a disc sample ($k_{p}$)
and the longitudinal vibration of a plate/bar sample ($k_{31}$) cannot
be compared directly based on vibration velocity, as detailed in \cite{Ural2010}. 

Comparing material properties based on vibration velocity seems natural,
as it an easy to recognize its significance, and materials properties
measured on samples having different dimensions can be compared. However,
comparing the $Q_{m}$ performance of material systems simply based
on vibration velocity is not a complete solution. The first reason
is already mentioned: the disc and the plate sample cannot be compared.
This will be explained in greater detail in the next section, but
to summarize the reason, the vibration velocity distribution profiles
have different shapes. The second reason is that vibration velocity,
although is not dependent on the size of the sample, is dependent
on the stiffness or the density of the sample. It is simple to tell
that a material with a higher mass density must have a higher mechanical
energy to reach the same vibration velocity than a material with a
lower mass density. However, simply judging from a vibration velocity
perspective, both situations are equal. 

The first goal of this paper is to prove analytically that vibration
velocity alone may result in misleading conclusions when directly
used to compare between material systems. Rather a new measure of
excitation, mechanical energy density should be used. This methodology
is demonstrated using published data from different material compositions
in the literature. The second goal of this work is to show that energy
density can be effectively used to determine anisotropic properties
in high power conditions by providing a similar comparison metric
between different piezoelectric resonators used to determine properties.

\section{Theoretical Treatment}

In this section, the non-dimensional quantity, mechanical energy density,
will be presented to normalize the dependence of $Q$ on material
vibration performance. The mechanical energy will be expressed in
terms of the maximum kinetic energy instead of the maximum elastic
energy because the compliance of the materials undergo nonlinearities,
whereas the mass density does not, thereby simplifying the calculations
for practical measurements.
\begin{table}
\caption{Parameter definitions\label{tab:Parameter-definitions}}

\centering{}%
\begin{tabular}{ccc}
\hline 
Symbol & Description & Units\tabularnewline
\hline 
$A$ & cross-sectional area of the plate & $\mathrm{m^{2}}$\tabularnewline
$\rho$ & mass density & $\mathrm{kg/m^{3}}$\tabularnewline
$r(x,t)$ & vibration distribution & $\mathrm{m/s}$\tabularnewline
$V_{RMS}$ & RMS tip vibration velocity  & $\mathrm{m/s}$\tabularnewline
$n$ & mode number for plate ($k_{31}$ mode) & $-$\tabularnewline
$f$ & frequency & $\mathrm{1/s}$\tabularnewline
$U_{e,31}$ & energy of a plate oscillator & $\mathrm{J}$\tabularnewline
$U_{e,p}$ & energy of a disc oscillator & $\mathrm{J}$\tabularnewline
$u_{e,31}$ & energy density of a plate oscillator & $\mathrm{J/m^{3}}$\tabularnewline
$u_{e,p}$ & energy density of a disc oscillator & $\mathrm{J/m^{3}}$\tabularnewline
$\theta$ & angle & $\mathrm{rad}$\tabularnewline
$m$ & mode number for disc ($k_{p}$ mode) & \tabularnewline
\hline 
\end{tabular}
\end{table}

\begin{figure}
\begin{centering}
\includegraphics[width=8cm]{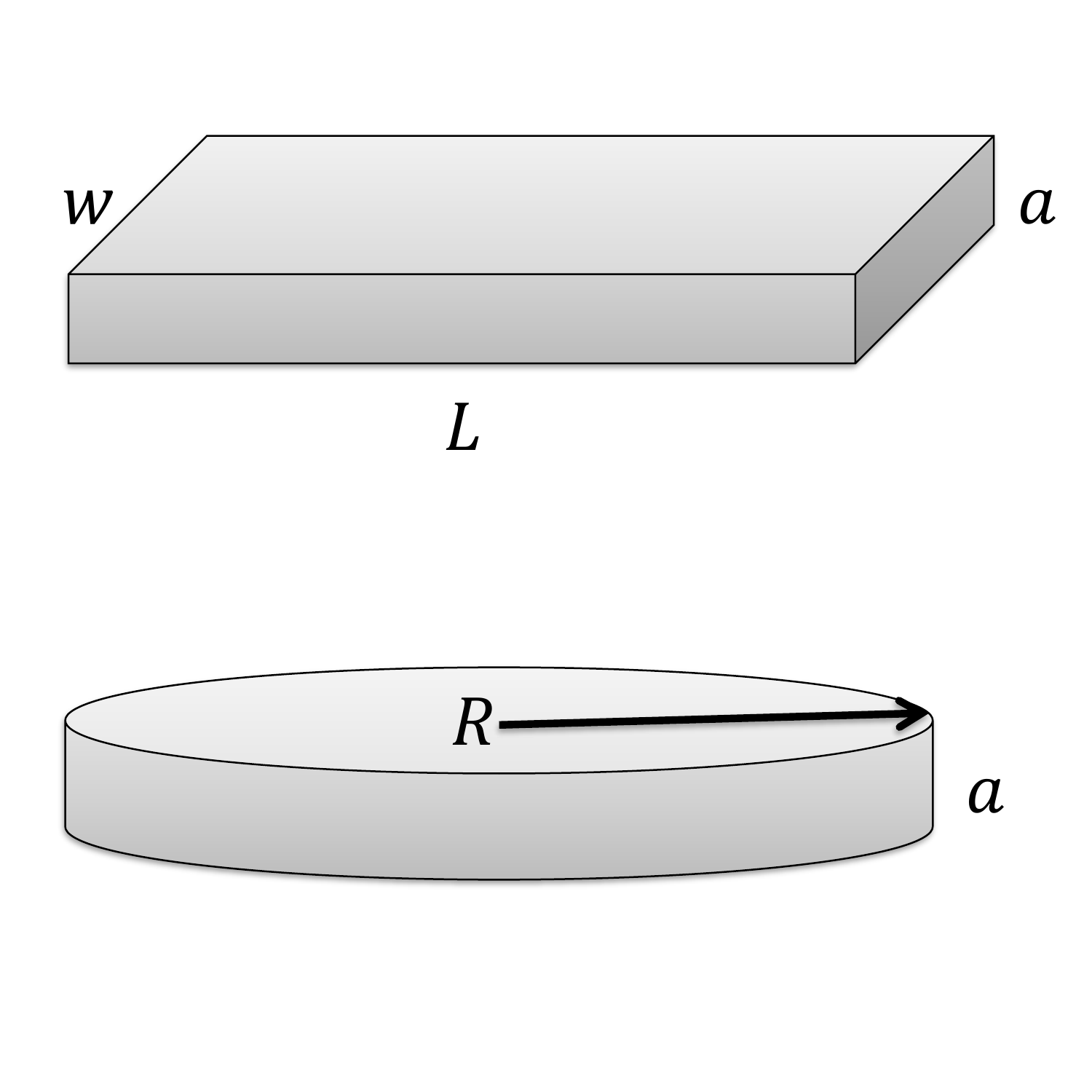}
\par\end{centering}

\caption{Geometry of plate and disc\label{fig:Geometry-of-plate}}
\end{figure}

Parameters used in the theoretical derivations are summarized in Table~
\ref{tab:Parameter-definitions}. For the $k_{31}$ mode, using the
geometry shown in Fig.~\ref{fig:Geometry-of-plate}, the mechanical
energy can be defined as the maximum kinetic energy, which is defined
by the maximum vibration velocity distribution $r_{31,max}(x)$ \cite{Soedel2004}

\begin{eqnarray}
U_{e,31} & = & \frac{1}{2}A\int\limits _{-\frac{L}{2}}^{\frac{L}{2}}\rho\left(r_{31,max}(x)\right)^{2}\,\mathrm{dx}.\label{eq:kinetic energy-1-2}
\end{eqnarray}

Assuming sinusoidal forcing at a frequency near the resonance frequency,
the spatial vibration can be described as

\begin{equation}
r_{31}(x,t)=V_{RMS}\sqrt{2}\sin\left(n\frac{\pi x}{L}\right)\sin\left(2\pi f\,t\right)
\end{equation}
and so the maximum velocity distribution is 
\begin{equation}
r_{31,max}(x)=V_{RMS}\sqrt{2}\sin\left(\frac{n\pi x}{L}\right).
\end{equation}
 The mechanical energy can be defined as

\begin{eqnarray}
U_{e,31} & = & \frac{1}{2}A\int\limits _{-\frac{L}{2}}^{\frac{L}{2}}\rho\left(V_{RMS}\sqrt{2}\sin\left(\frac{n\pi x}{L}\right)\right)^{2}\,dx\\
 & = & V_{RMS}^{2}2\rho\frac{1}{2}A\int\limits _{-\frac{L}{2}}^{\frac{L}{2}}\sin^{2}\left(\frac{n\pi x}{L}\right)\,dx\\
 & = & V_{RMS}^{2}\rho A\frac{L}{2}.\label{eq:mech eng plate-1}
\end{eqnarray}
Dividing by the volume, the energy density of the ceramic is found 

\begin{equation}
u_{e,31}=\frac{1}{2}\rho V_{RMS}^{2}.\label{eq:plate energy density-1}
\end{equation}
Notice that the mechanical energy density does not depend on mode
number nor on the geometrical dimensions, but it does depend on the
mass density and the vibration velocity. It is proposed that this
quantity should be compared between material compositions because
this will be the value which will be most useful for applications. 

For the $k_{p}$ mode (radial vibration), the mechanical energy is
described by the the maximum kinetic energy through the maximum vibration
velocity distribution, $r_{p,max}(x,t)$, is \cite{Soedel2004}

\begin{eqnarray}
U_{e,d}(x,t) & = & \frac{1}{2}a\int\limits _{0}^{R}\int\limits _{0}^{2\pi}\rho\left(r_{p,max}(x)\right)^{2}x\,\mathrm{d\theta dx}\label{eq:kinetic energy-1-1-1}
\end{eqnarray}
Given that the plate is uniform with respect to the angular direction
for the $k_{p}$ mode, the integral with respect to $\theta$ can
be evaluated

\begin{equation}
U_{e,d}=\pi\rho a\int\limits _{0}^{R}\left(r_{p,max}(x)\right)^{2}x\,\mathrm{dx.}
\end{equation}
The mode shape (vibration velocity profile) for radial vibration
of a disc is given by

\begin{equation}
r_{p}(x)=J_{1}(\lambda_{m}x)
\end{equation}
\begin{table}
\caption{Eigenvalue solutions for radially vibrating disc having free boundary
conditions\label{tab:Eigenvalue-solutions-for-1}}

\centering{}%
\begin{tabular}{cc}
\hline 
$m$ & $(\lambda R)_{m}$\tabularnewline
\hline 
1 & 2.049\tabularnewline
2 & 5.389\tabularnewline
3 & 8.572\tabularnewline
\hline 
\end{tabular}
\end{table}
\begin{figure}
\begin{centering}
\includegraphics[width=10cm]{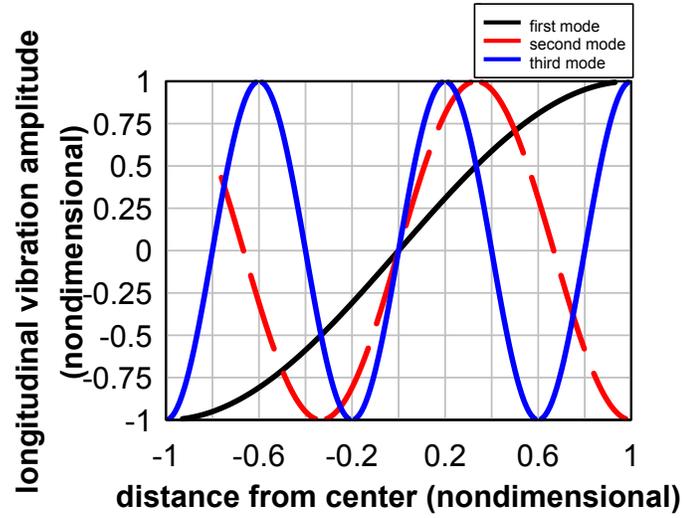}
\par\end{centering}

\caption{Mode shapes of vibration for longitudinal vibration ($k_{31},\,k_{33},\,\mathrm{and}\,k_{t}$)\label{fig:Mode-shapes-of-long}}
\end{figure}
\begin{figure}
\begin{centering}
\includegraphics[width=10cm]{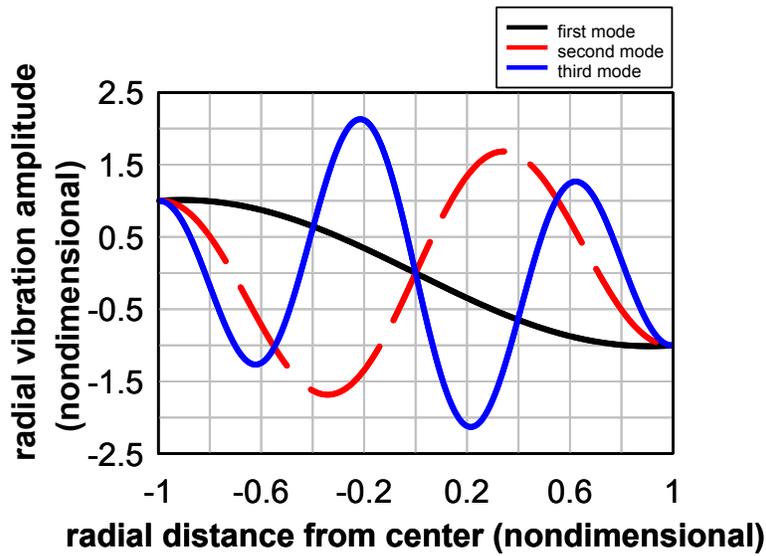}
\par\end{centering}

\caption{Mode shapes of vibration for radial vibration ($k_{p}$)\label{fig:Mode-shapes-of-radial}}
\end{figure}
where, $J_{1}$ is the Bessel function of the first kind, $\lambda_{m}=\frac{(\lambda R)_{m}}{R}$,
and values of $(\lambda R)_{m}$ are shown in Table \ref{tab:Eigenvalue-solutions-for-1}.
Normalizing the mode shape to the tip vibration velocity,

\begin{equation}
r_{p,max}(x)=V_{RMS}\sqrt{2}\frac{J_{1}(\lambda_{m}x)}{J_{1}(\lambda_{m}R)}=V_{RMS}\sqrt{2}\frac{J_{1}(\lambda_{m}x)}{J_{1}((\lambda R)_{m})}.
\end{equation}
The energy of the disc oscillator can be described as 

\begin{equation}
U_{e,p}=\pi\rho a\int\limits _{0}^{R}\left(V_{RMS}\sqrt{2}\frac{J_{1}(\lambda_{m}x)}{J_{1}((\lambda R)_{m})}\right)^{2}x\,\mathrm{dx}.
\end{equation}
By dividing by the volume, the mechanical energy density is described
as

\begin{equation}
u_{e,p}=\frac{2\rho V_{RMS}^{2}\int\limits _{0}^{R}\left(\frac{J_{1}(\lambda_{m}x)}{J_{1}((\lambda R)_{m})}\right)^{2}x\,\mathrm{dx}}{R^{2}}.
\end{equation}
Solving the integral, the energy density for the first three modes
can be expressed as

\begin{equation}
u_{e,p,1}=0.783\rho V_{RMS}^{2}\label{eq: disk energy density-1}
\end{equation}

\begin{equation}
u_{e,p,2}=0.969\rho V_{RMS}^{2}
\end{equation}

\begin{equation}
u_{e,p,3}=0.989\rho V_{RMS}^{2}
\end{equation}

The mechanical energy density increases with the vibration mode in
discs because as the mode number increases the point of maximum vibration
velocity does not occur at the tip velocity, although there is a local
maximum of the vibration velocity at the edge of the sample due to
the mechanical condition that the stress must be zero at the boundaries.
This is different from the case of longitudinal vibration, where the
edge vibration level is equal to or greater than the vibration occurring
within the structure. The vibration velocity distribution (mode shape)
of the longitudinal type and radial type vibration for the first three
modes of vibration can be found in Figs. \ref{fig:Mode-shapes-of-long}
and \ref{fig:Mode-shapes-of-radial}. Thus, normalizing the vibration
velocity shape to the edge of the disc increases mechanical energy
of the system. Practically, a lower vibration velocity is necessary
to achieve a higher mechanical energy density in the case of the disc
(radial vibration) in comparison to the plate (longitudinal vibration).
Considering the first mode of vibration, the ratio between the mechanical
energy density of plates and discs is

\begin{equation}
\frac{u_{p}}{u_{31}}=1.57.
\end{equation}
The edge vibration velocity in the plate and the edge vibration velocity
in the disc which brings about the same mechanical energy density,
assuming the mass density and volume is the same, is

\begin{equation}
\frac{V_{RMS,31}}{V_{RMS,p}}=1.252.
\end{equation}

It is proposed here that researchers use energy density when reporting
material property values because energy density is a common comparison,
both between different modes of vibration and between different compositions
which have different mass densities (such as lead free and lead containing
piezoceramics). Because different tensor properties contribute toward
plate and disc vibration, the quality factor according to different
tensor elastic properties are not theoretically equal. However, using
mechanical energy density, the quality factor data measured from plate
and disc samples will have a fair comparison. The derivation presented
solves for the energy density as a function of vibration velocity,
which results in the energy density being dependent on the mass density.
The mechanical energy density can also be solved in terms of strain,
in which case the energy density expression would depend on elastic
compliance. This approach is not taken because the compliance has
nonlinearity, whereas mass density is stable.

\section{Analysis of $Q_{m}$ according to mechanical energy density}

\subsection{Mechanical Energy Density as a Figure of Merit}

Vibration velocity is commonly used as a figure of merit when comparing
high power capabilities of piezoelectric materials. Furthermore, it
can be proven that the strain level for samples with the same vibration
velocity is the identical, allowing comparison of data between samples
having different dimensions. However, this is true provided that the
mode of vibration is the same. Also, if the material systems are different,
the mass density can be significantly different, which means that
the mechanical energy density of the compared materials will not be
the same for a common vibration level. This can be clearly seen in
Eqs. \ref{eq:plate energy density-1} and \ref{eq: disk energy density-1}
where the mechanical energy density is proportional to the mass density. 

The authors of this paper propose that mechanical energy density should
be used as a common comparison condition for high power properties
instead of solely using vibration velocity. This is because the vibration
velocity does not account for the effects of vibration mode type and
material density. By utilizing the mechanical energy density approach,
one will be in effect normalizing the vibration velocity such that
the materials can be suitably compared. In most applications, the
piezoelectric materials will be working against an external load,
be it a resistive load for a piezoelectric transformer, or a physical
load for an ultrasonic motor. Such conditions are imposing on the
material to work on external environments, in which cases energy density
is a better figure of merit than vibration velocity.

\subsection{Analysis of High Power Behavior by using Mechanical Energy Density
as a Figure of Merit}

In this portion, high power behavior of three distinctly different
piezoelectric materials (i.e. Bi-based, alkaline-based, lead-based)
will be analyzed using the new mechanical energy density approach.
For the purposes of comparison, results from BNKLT based ($\mathrm{0.88(Bi_{0.5}Na_{0.5})TiO_{3}}$
$\mathrm{-0.04(Bi_{0.5}Li_{0.5})TiO_{3}}$ $\mathrm{-0.08(Bi_{0.5}Na_{0.5})TiO_{3}+}\mathrm{MnCO_{3}\,0.4\,wt}$\%)\cite{Nagata2010},
KN based $\mathrm{(KNbO_{3}+MnCO_{3}\,0.8\,wt}$\%)\cite{Nagata2010},
BNT\textendash BT\textendash BNMN based ($\mathrm{0.82(Bi_{0.5}Na_{0.5})TiO_{3}}$
$\mathrm{\text{\textendash}0.15BaTiO_{3}}$ $\mathrm{\text{\textendash}0.03(Bi_{0.5}Na_{0.5})}$
$\mathrm{(Mn_{1/3}Nb_{2/3})O_{3}}$)\cite{Tou2009}, commercial hard
PZT ($\mathrm{Pb(Zr,Ti)O_{3})}$\cite{Nagata2010}, and NKN based
($\mathrm{(Na_{0.5}K_{0.5})}$ $\mathrm{(Nb_{0.97}Sb_{0.03})O_{3}}$
$\mathrm{+CuO\,1.5\,
}$\%) \cite{Gurdal2013} ceramics found in the literature were used.
The samples used in these studies feature longitudinal vibration of
a plate shape ($k_{31}$), except for NKN \cite{Gurdal2013}, where
a disc sample was used having radial vibration ($k_{p}$). The properties
at low level excitation for the various materials can be found in
Tab. \ref{tab:Densities-of-materials}. Regarding high power measurements,
the studies from \cite{Nagata2010} and \cite{Tou2009} used the burst/transient
method to determine the mechanical quality factor according to vibration
velocity. In the burst method, the sample receives large field excitation
for a set number of cycles, after which the excitation is removed
and the vibration is allowed to decay. By measuring the rate of decay
using current measurement, the mechanical quality factor can be calculated
\cite{Umeda1998}. Gurdal et al. measured the mechanical quality factor
under continuous excitation at the given vibration velocity\cite{Gurdal2013}
. This was accomplished by controlling the voltage input while sweeping
the frequency across the resonance frequency while logging impedance.
Using the 3dB bandwidth of the impedance around the resonance frequency,
the mechanical quality factor can be calculated. 

\begin{figure}
\begin{centering}
\includegraphics[width=10cm]{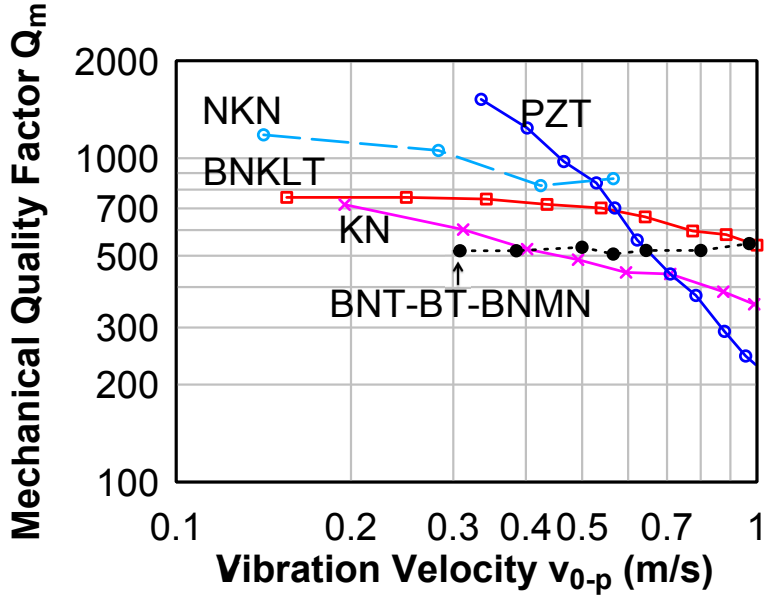}
\par\end{centering}

\caption{Change in mechanical factor with increasing vibration level\label{fig:Change-in-mechanical vibration}}
\end{figure}

Figure \ref{fig:Change-in-mechanical vibration} shows the change
in the mechanical quality factor as a result of change in the peak
vibration velocity. This figure shows that the quality factor tends
to degrade with increasing vibration levels. The degree to which each
material degrades is associated with onset of domain wall motion,
but that is not the point of discussion of this paper. With increasing
vibration levels the lead-free material systems seem to maintain their
low power $Q_{m}$ values with minimal degradation, while the $Q_{m}$
of PZT degrades very quickly. From this figure, it is apparent that
all the Pb-free materials shown exceed the quality factor of the PZT
sample when the vibration velocity is large enough. BNKLT exceeds
the quality factor of PZT at 0.57 m/s, BNT-BT-BNMN at 0.64 m/s, KN
at 0.71 m/s, and NKN at 0.53 m/s.

\begin{figure}
\begin{centering}
\includegraphics[clip,width=10cm]{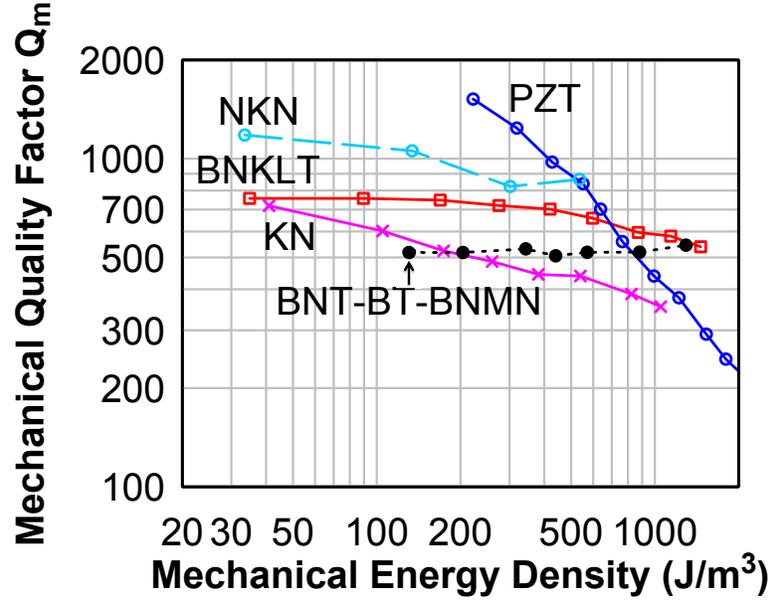}
\par\end{centering}

\caption{Change in mechanical quality factor with mechanical energy density\label{fig:Change-in-Mechanical energy density}}
\end{figure}

\begin{table*}
\begin{centering}
\begin{tabular}{cccccc}
\hline 
Material & $d_{31}\mathrm{(pC/N)}$ & $s_{11}^{E}$$\mathrm{(pm^{2}/N)}$ & $Q_{m,31}$ & $\rho(kg/m^{3})$ & Reference\tabularnewline
\hline 
BNKLT & 20 & 7.6 & 730 & 5800 & \cite{Nagata2010}\tabularnewline
KN & 25 & 7.7 & 820 & 4300 & \cite{Nagata2010}\tabularnewline
NKN & 99 ($d_{33}$) & - & 1051 $(Q_{m,p})$ & 4280 & \cite{Gurdal2013}\tabularnewline
Hard PZT & 125 & 12.1 & 1800  & 7900 & \cite{Nagata2010}\tabularnewline
BNT\textendash BT\textendash BNMN & 110 ($d_{33}$) & 8.8$(s_{33}^{E})$ & 500$(Q_{m,33})$ & 5500 & \cite{Tou2009}\tabularnewline
\hline 
\end{tabular}
\par\end{centering}

\caption{Material properties under low level excitation\label{tab:Densities-of-materials}}
\end{table*}

Figure \ref{fig:Change-in-Mechanical energy density} shows the change
in the mechanical quality factor as a change in the mechanical energy
density. The end of each of the curves has a tip vibration velocity
of 1 m/s, except NKN whose is 0.57 m/s. Using mechanical energy density,
the vibration velocity for each system is effectively normalized with
its density and vibration mode. Looking at this figure, it is apparent
that the quality factor of KN no longer exceeds that of PZT for any
mechanical energy condition. This is because after taking into consideration
the low mass density of KN, its performance becomes less significant.
BNT-BT-BNMN exceeds the quality factor of PZT at a mechanical energy
condition of 850 $\mathrm{J/m^{3}}$. This corresponds to the vibration
velocity condition of 0.76 m/s for BNT-BT-BNMN and 0.64 m/s for PZT.
BNKLT exceeds the quality factor of PZT for mechanical energy densities
of higher than 700 $\mathrm{J/m^{3}}$, which corresponds to a vibration
condition of 0.64 m/s for BNKLT and 0.55 m/s for PZT, which is different
from the conclusion which could be made from Figure \ref{fig:Change-in-mechanical vibration},
where the $Q_{m}$ of BNKLT exceeded that of PZT at 0.5 m/s. NKN exceeded
the quality factor of the PZT at 540 $\mathrm{J/m^{3}}$, which is
equivalent to 0.56 m/s for NKN and 0.53 m/s for PZT. 

The general trend observed is that the quality factor of the lead-free
materials does not suffer from the sharp degradation as the commercial
PZT sample does. As a result, although the mechanical quality factors
of the lead-free materials are lower at low vibration velocities,
they are higher past a sufficiently high mechanical excitation level.
For all of the $k_{31}$ samples, the equivalent mechanical energy
density condition resulted in a larger vibration velocity for the
lighter lead-free sample rather than the more dense PZT. Because the
energy density takes into account mass density when calculating the
performance level, the performance of PZT is apparently enhanced because
of its large density. A larger vibration velocity is required from
the lighter lead-free materials to reach the equivalent vibration
velocity for the heavier PZT. With regards to NKN, the $k_{p}$ sample,
the equivalent mechanical energy density condition yields an almost
identical vibration velocity for the PZT and the NKN material system.
This is because, the lighter density of NKN reduces its relative performance
to the more dense PZT, but the mode of vibration of the NKN sample
($k_{p}$) increases its performance. That being said, the density
or the vibration mode does not actually enhance or the increase the
degradation characteristics of the material. In fact, the mechanical
energy density approach presented allows one to remove their effects
and to compare material properties in a better justified manner. 

The purpose of this section was to demonstrate the observations made
comparing $Q_{m}$ performance as a function of vibration velocity
and as a function of mechanical energy density will be different.
Comparison of the $Q_{m}$ degradation of materials using vibration
velocity as the common condition has limitations, as was discussed.
Although general trends seen using vibration velocity as the figure
of merit still hold true when observing the mechanical quality factor
with regard to mechanical energy density, relative material performance
can best be evaluated when utilizing mechanical energy density as
the common comparison condition.

\subsection{Comparison of $Q_{A}$ in $k_{31}$ and $k_{p}$ resonators through
energy density}

\begin{figure}
\begin{centering}
\subfloat[]{\begin{centering}
\includegraphics[height=7cm]{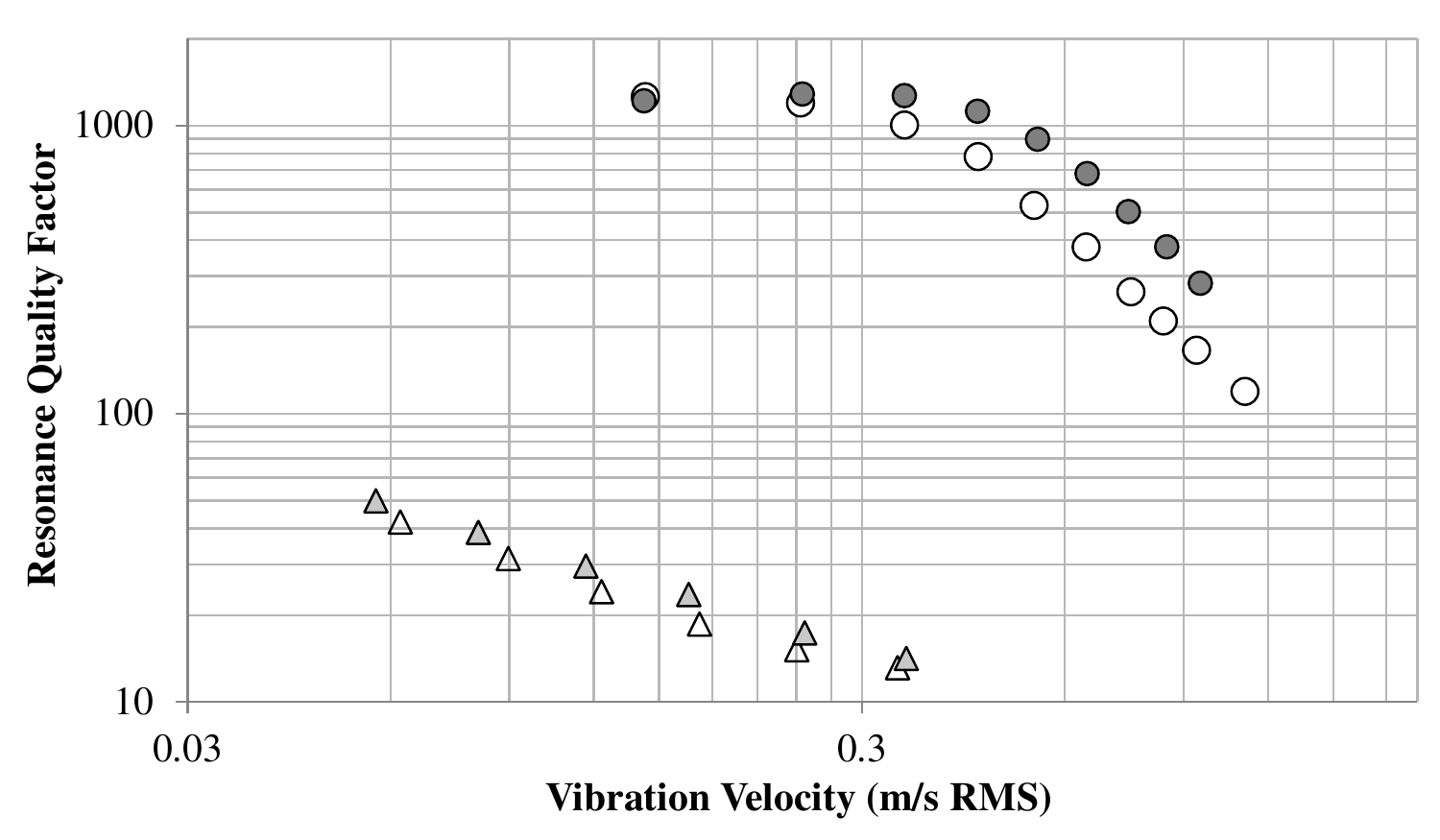}
\par\end{centering}

}
\par\end{centering}

\centering{}\medskip{}
\subfloat[]{\begin{centering}
\includegraphics[height=7cm]{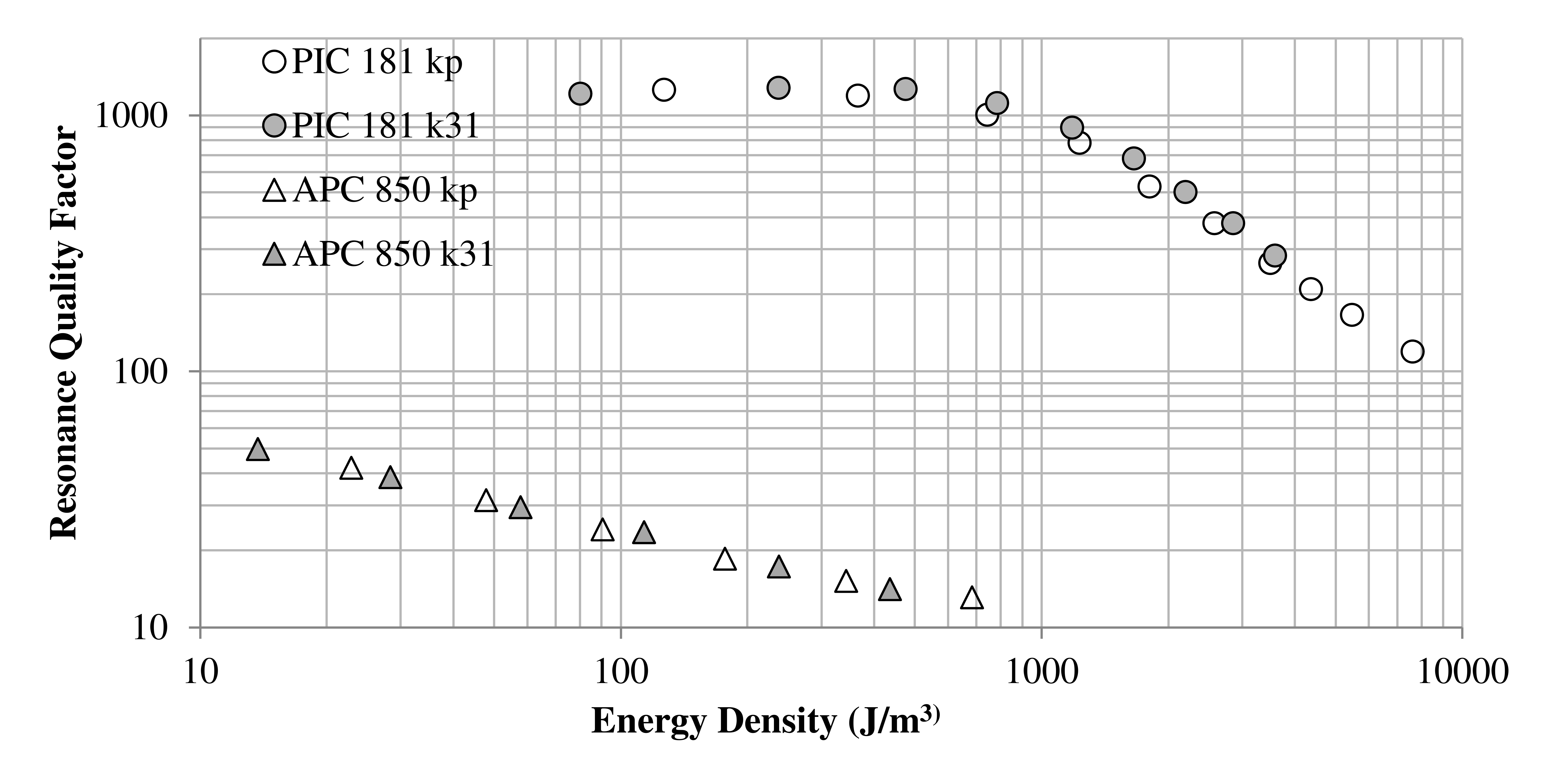}
\par\end{centering}

}\caption{Comparison of the resonance quality factor $Q_{A}$ versus vibration
velocity (a) and energy density (b) in hard (PIC 181) and soft PZT
(APC 850) \label{fig: energy density} }
\end{figure}

The energy density versus $Q_{A}$ is plotted in Fig.~\ref{fig: energy density}
for hard and soft PZT of $k_{31}$ and $k_{p}$ resonator geometries.
This data was taken using the burst method operated at antiresonance.
After accounting for the energy density of each mode type ($k_{31}$
and $k_{p}$) and the mass density of the material, the quality factors
of both modes agree, showing they are comparable between different
modes. Previously, it may have seemed as though the $k_{p}$ degrades
quickly; however, this is due to the increased energy density stored
per unit of vibration velocity. That being said, the $k_{p}$ mode's
energy potential is omnidirectional in the plane of vibration, which
may make it difficult to apply this resonator in devices. However,
in applications such as structural health monitoring and piezoelectric
transformers, this feature can be effectively utilized, in which case
it is important to be able to compare the modes. Anisotropy is a fundamental
quality of piezoelectric materials, which implies that losses are
also anisotropic. Thus, although the quality factor for the $k_{31}$
and $k_{p}$ modes in this analysis were of similar levels, it is
not possible to assume this generalize to all crystalline piezoelectric
ceramics.

\section{Conclusion}

Analysis of the mechanical quality factor ($Q_{m}$) as a function
of vibration velocity is typically used to compare the high power
behavior of piezoelectric materials. Using vibration velocity as a
figure of merit might be misleading if distinctly different materials
and vibration modes are compared. In this paper, a new figure of merit
is introduced to compare high power behavior of different piezoelectric
materials: mechanical energy density. The formulas describing the
mechanical energy density were solved for longitudinal vibration ($k_{31},\,k_{33},\,\mathrm{and}\,k_{t}$)
and radial vibration ($k_{p}$) type transducers. The new analysis
method was applied to data regarding lead-free (i.e. KN, BNKLT, BNT-BT-BNMN,
NKN) and lead-based (hard PZT) piezoelectric materials found in the
literature. Analyzing mechanical quality factor as a function mechanical
energy density gave different conclusions than analyzing it with regard
to vibration velocity. When analyzing $Q_{m}$ behavior versus vibration
velocity, all lead-free ceramics surpassed hard-PZT. However, when
analyzing $Q_{m}$ behavior versus mechanical energy lead-free ceramics
$Q_{m}$ the trends observed differed. As a result, comparing the
high power of behavior of different materials with mechanical energy
density provides a better standard of comparison between different
materials and different sample geometries than vibration velocity.
This is because mechanical energy density accounts for the mass density
of the material system and also accounts for type of vibration, thereby
normalizing vibration velocity to mechanical energy density. The mechanical
energy density approach is necessary to compare mechanical loss in
piezoelectric ceramics for different material systems and vibration
modes. Finally, the energy density concept was applied to hard and
soft PZT materials of $k_{p}$ and $k_{31}$ resonator geometries.

\section*{Acknowledgments}

The authors would like to acknowledge the support by funding from
The Office of Naval Research (ONR) Grant Number: ONR N00014-12-1-1044.

\bibliographystyle{ieeetr}

\pagebreak{}

\end{document}